\documentclass{emulateapj}
\def \be{\begin{equation}}
\def \ee{\end{equation}}
\def \bdm{\begin{eqnarray}}
\def \edm{\end{eqnarray}}
\begin{document}
\title{Time-Dependent Perpendicular Transport of Energetic Particles for Different Turbulence Configurations and Parallel Transport Models}
\author{J. Lasuik and A. Shalchi}
\affil{Department of Physics and Astronomy, University of Manitoba, Winnipeg, Manitoba R3T 2N2, Canada}
\email{andreasm4@yahoo.com}
\begin{abstract}
Recently a new theory for the transport of energetic particles across a mean magnetic field was presented.
Compared to other non-linear theories the new approach has the advantage that it provides a full time-dependent
description of the transport. Furthermore, a diffusion approximation is no longer part of that theory.
It is the purpose of the current paper to combine this new approach with a time-dependent model for
parallel transport and different turbulence configurations in order to explore the parameter regimes
for which we get ballistic transport, compound sub-diffusion, and normal Markovian diffusion.
\end{abstract}
\section{Introduction}
The transport of electrically charged particles in a turbulent magnetized plasma is a topic of great interest in modern
physics. Especially in astrophysics scientists are keen on understanding the motion of cosmic rays and solar energetic
particles through the universe (see, e.g., Schlickeiser 2002 and Zank 2014 for reviews). If such particles would only interact
with a constant magnetic field, their trajectory would be a perfect helix. However, real particles experience scattering
due to the interaction with turbulent electric and magnetic fields. Therefore, one finds different transport processes such
as parallel diffusion or stochastic acceleration. In particular the motion of energetic particles across the mean magnetic
field was the subject of numerous theoretical studies (see, e.g., Shalchi 2009 for a review).

A simple analytical description of perpendicular transport is provided by quasi-linear theory (see Jokipii 1966 for the
original presentation of this approach) where it is assumed that perpendicular diffusion is caused by particles following
magnetic fields lines which themselves behave diffusively. Characteristic for this type of transport is that the corresponding
perpendicular mean free path does not depend on particle energy nor does it depend on other particle properties. It is entirely
controlled by magnetic field parameters. This type of transport is often referred to as the {\it field line random walk (FLRW)}
limit. More realistic descriptions have been developed later. It was shown, for instance, that parallel diffusion suppresses
perpendicular transport to a sub-diffusive level. Analytical descriptions of this type are usually called {\it compound sub-diffusion}
(see, e.g., K\'ota \& Jokipii 2000 or Webb et al. 2006).

Comprehensive numerical studies of perpendicular transport have been performed showing that one can indeed find sub-diffusive
transport if a turbulence configuration without any transverse structure in considered (see, e.g., Qin et al. 2002a). However,
if there is transverse complexity of the turbulence, diffusion is restored (see, e.g., Qin et al. 2002b). Matthaeus et al. (2003)
and Qin et al. (2002b), therefore, distinguished between first diffusion (meaning quasi-linear transport) and second diffusion.
The latter process must be described by some type of non-linear interaction between particles and magnetic fields and it must
also be related to the transverse structure of the turbulence. It has to be emphasized that the recovery of diffusion due to
collisions was described in the famous work of Rechester \& Rosenbluth (1978) but Coulomb collisions, albeit relevant in
laboratory plasmas, should not be important in astrophysical scenarios such as the solar wind or the interstellar medium.

Different attempts to describe second diffusion have been presented in the past such as the pioneering work of Matthaeus et al. (2003).
A few years later the so-called {\it unified non-linear transport (UNLT) theory} has been derived (see Shalchi 2010) which contains
the field line diffusion theory of Matthaeus et al. (1995), quasi-linear theory, as well as a Rechester \& Rosenbluth type of diffusion
as special limits (see Shalchi 2015 for a detailed discussion of this matter). However, the aforementioned theories rely on a diffusion
approximation together with a late time limit. Therefore, such theories do not describe the early ballistic motion of the particles,
nor do they explain the sub-diffusive regime and the recovery of diffusion. In Shalchi (2017) a time-dependent version of UNLT theory
was developed. Although this description still relies on approximations and assumptions such as {\it Corrsin's independence hypothesis}
(see Corrsin 1959), one is now able to describe the transport as a full time-dependent process. Furthermore, the latter theory provides
a simple condition which needs to be satisfied in order to find second diffusion, namely $\langle ( \Delta x )^2 \rangle \gg 2 \ell_{\perp}^2$.
Here we have used the mean square displacement of possible particle orbits as well as a characteristic length scale of the turbulence
in the perpendicular direction. Before this condition is satisfied, perpendicular transport is either ballistic, quasi-linear,
or sub-diffusive.

It is the purpose of the current paper to present a detailed analytical description of time-dependent perpendicular transport.
We employ a general model for the transport of particles in the parallel direction containing ballistic and diffusive regimes.
Furthermore, we employ different analytical models for magnetic turbulence such as the slab model, a noisy slab model, a
Gaussian model, the two-dimensional model, as well as a two-component turbulence model. In all cases we compute the running
perpendicular diffusion coefficient as a function of time to explore the different transport regimes.
\section{Full time-dependent description of perpendicular transport}
A time-dependent theory for perpendicular transport was developed in Shalchi (2017) based on ideas discussed
in Matthaeus et al. (2003) and Shalchi (2010). By using guiding center coordinates instead of particle coordinates,
{\it Corrsin's independence hypothesis}, and by assuming that the averages over particles properties can be written
as product of parallel and perpendicular correlation functions, the following equation was derived
\be
\langle V_x (t) V_x (0) \rangle = \frac{1}{B_0^2} \int d^3 k \; P_{xx} \bigl( \vec{k}, t \bigr) \xi \left( k_{\parallel}, t \right)
e^{- \frac{1}{2} \langle ( \Delta x )^2 \rangle k_{\perp}^2}
\label{centraleq}
\ee
where we have used the $x$-component of the guiding center velocity $V_x (t)$, the mean magnetic field $B_0$,
the $xx$-component of the magnetic correlation tensor $P_{xx}$ (see below for some examples), and the mean square displacement
in the $x$-direction $\langle ( \Delta x )^2 \rangle$. Eq. (\ref{centraleq}) is valid for axi-symmetric and dynamical
turbulence. However, in the current paper we only consider magnetostatic turbulence where by definition
$P_{xx} (\vec{k},t) = P_{xx} (\vec{k})$. Furthermore, Eq. (\ref{centraleq}) is based on the assumption the
$\delta B_z \ll B_0$. If one considers isotropic turbulence, for instance, one has to ensure that the turbulent
magnetic field is not too strong. To generalize Eq. (\ref{centraleq}) to allow turbulence which is not axi-symmetric would
be straightforward. In this case Eq. (\ref{centraleq}) would be replaced by a set of four coupled ordinary differential
equations.

In Eq. (\ref{centraleq}) we have also used $\xi (k_{\parallel},t) = \xi (k_{\parallel},t_1=t,t_2=0)$ with the parallel
correlation function
\bdm
\xi \left( k_{\parallel}, t_1, t_2 \right)
& = & k_{\parallel}^{-2} \left< \left( \frac{d}{d t_1} e^{i z (t_1) k_{\parallel}} \right)
\left( \frac{d}{d t_2} e^{- i z (t_2) k_{\parallel}} \right) \right> \nonumber\\
& = & k_{\parallel}^{-2} \frac{d}{d t_1} \frac{d}{d t_2}
\left< e^{i \left[ z \left( t_1 \right) - z \left( t_2 \right) \right] k_{\parallel}} \right>.
\label{definexi}
\edm
The latter function was explored in Shalchi et al. (2011) based on the cosmic ray Fokker-Planck equation
\be
\frac{\partial f}{\partial t} + v \mu \frac{\partial f}{\partial z}
= \frac{\partial}{\partial \mu} \left[ D_{\mu\mu} \left( \mu \right) \frac{\partial f}{\partial \mu} \right]
\label{FPeq}
\ee
with isotropic pitch-angle scattering coefficient $D_{\mu\mu} = (1-\mu^2)D$. A detailed discussion of the analytical
form of $D_{\mu\mu}$ and the validity of the isotropic regime can be found in Shalchi et al. (2009). The solution of
Eq. (\ref{FPeq}) describes the parallel motion of energetic particles while they experience pitch-angle scattering.
As explained in Shalchi (2006), the solution describes the parallel motion as a ballistic motion at early times and
then the solution becomes diffusive. This is exactly what one observes in test-particle simulations performed in the
past (see, e.g., Qin et al. 2002a and Qin et al. 2002b). A general solution of the two-dimensional Fokker-Planck equation
is difficult to find. Within a two-dimensional sub-space approximation, it was derived in Shalchi et al. (2011) that
\be
\xi \left( k_{\parallel}, t \right) = \frac{v^2}{3} \frac{1}{\omega_+ - \omega_-}
\left[ \omega_+ e^{\omega_+ t} - \omega_- e^{\omega_- t} \right]
\label{formforW}
\ee
with the parameters
\bdm
\omega_{\pm} & = & - D \pm \sqrt{D^2 - \frac{1}{3} \left( v k_{\parallel} \right)^2} \nonumber\\
& = & - \frac{v}{2 \lambda_{\parallel}} \pm \sqrt{ \left( \frac{v}{2 \lambda_{\parallel}} \right)^2 - \frac{1}{3} \left( v k_{\parallel} \right)^2}
\label{omegapm}
\edm
where we have used the parallel mean free path $\lambda_{\parallel} = v /(2 D)$.

In order to compute a time-dependent or running diffusion coefficient
\be
d_{\perp} (t) = \frac{1}{2} \frac{d}{d t} \langle (\Delta x)^2 \rangle,
\label{rundiff}
\ee
one has to employ the {\it TGK (Taylor-Green-Kubo) formulation} (see Taylor 1922, Green 1951, and Kubo 1957)
\be
d_{\perp} (t) = \Re \int_{0}^{t} d t' \; \left< V_x (t') V_x (0) \right>
\label{tgk}
\ee
where we assumed that the initial diffusion coefficient is zero $d_{\perp} (0) = 0$. The TGK formula can easily be combined
with  Eq. (\ref{centraleq}) yielding
\be
\frac{d^2}{d t^2} \langle (\Delta x)^2 \rangle
= \frac{2}{B_0^2} \int d^3 k \; P_{xx} \bigl( \vec{k} \bigr)
\xi \left( k_{\parallel}, t \right) e^{- \frac{1}{2} \langle ( \Delta x )^2 \rangle k_{\perp}^2}.
\label{centraleq2}
\ee
The latter ordinary differential equation can be evaluated numerically for any given turbulence model described by the
tensor component $P_{xx}$. After obtaining the second moment $\langle (\Delta x)^2 \rangle$, Eq. (\ref{rundiff}) can
be employed in order to compute the running diffusion coefficient $d_{\perp} (t)$. We call this approach time-dependent
UNLT theory. We like to emphasize that we only compute the second moment and use this to distinguish between normal
diffusion ($\langle (\Delta x)^2 \rangle \propto t$) and other types of transport such as compound sub-diffusion
where $\langle (\Delta x)^2 \rangle \propto t^{1/2}$. A more detailed analysis would provide other, higher, moments
and maybe even the full particle distribution function. The theory discussed here does not allow to compute such
quantities nor is this usually done in the field of transport theory (see again Qin et al. 2002a, Qin et al. 2002b,
and Matthaeus et al. 2003 for examples).
\section{Relation to Previous Results and Theories}
The main aim of the current paper is to present a detailed investigation of time-dependent perpendicular transport for a variety
of turbulence models. In the current section, however, we relate time-dependent UNLT theory to previous results such
as ballistic transport, compound sub-diffusion for slab turbulence, as well as UNLT theory based on the diffusion approximation.
\subsection{Early Times and Ballistic Transport}
For early times $t=0$ we have according to Eq. (\ref{formforW})
\be
\xi \left( t, 0 \right) = \frac{v^2}{3}
\ee
and, thus, Eq. (\ref{centraleq}) becomes
\be
\langle V_x (t) V_x (0) \rangle = \frac{v^2}{3 B_0^2} \int d^3 k \; P_{xx} \bigl( \vec{k} \bigr)
= \frac{v^2}{3 B_0^2} \delta B_x^2
\ee
where we have used $\langle (\Delta x)^2 \rangle_{t=0} = 0$ also. After integrating the latter formula over time,
and using Eq. (\ref{tgk}), we obtain for the running diffusion coefficient
\be
d_{\perp} (t) = \frac{v^2}{3} \frac{\delta B_x^2}{B_0^2} t.
\label{runningball}
\ee
If we integrate again, we find
\be
\langle \left( \Delta x \right)^2 \rangle = \frac{v^2}{3} \frac{\delta B_x^2}{B_0^2} t^2.
\ee
The motion obtained here corresponds to a ballistic motion where particles move unperturbed in the parallel direction 
while they follow ballistic magnetic field lines. We expect that for any given turbulence model, the running
diffusion coefficient can be approximated by Eq. (\ref{runningball}) if early enough times are considered.
\subsection{Slab Turbulence}
For slab turbulence, corresponding to turbulence without any transverse structure, we have by definition
\be
P_{nm}^{slab} (\vec{k}) = g^{slab} \left( k_{\parallel} \right) \frac{\delta (k_{\perp})}{k_{\perp}} \delta_{nm}
\label{defineslab}
\ee
for $n,m = x,y$. Because of the solenoidal constraint all other components of this tensor are zero. Due to the Dirac delta therein,
Eq. (\ref{centraleq}) becomes
\be
\frac{d^2}{d t^2} \langle (\Delta x)^2 \rangle
= \frac{8 \pi}{B_0^2} \int_{0}^{\infty} d k_{\parallel} \; g^{slab} \left( k_{\parallel} \right) \xi \left( k_{\parallel}, t \right).
\label{slabode}
\ee
Here we have used the one-dimensional spectrum of the slab modes $g^{slab} (k_{\parallel})$. If we combine the latter form
with the TGK formula (\ref{tgk}) and model (\ref{formforW}), we find after straightforward algebra
\be
d_{\perp} (t) = \frac{4 \pi v^2}{3 B_0^2} \int_{0}^{\infty} d k_{\parallel} \; g^{slab} \left( k_{\parallel} \right)
\frac{1}{\omega_+ - \omega_+} \nonumber\\
\left( e^{\omega_+ t} - e^{\omega_- t} \right)
\label{dperpslab}
\ee
which was originally derived in Shalchi et al. (2011). In the latter paper a more detailed discussion can be found
as well as a visualization of the running perpendicular diffusion coefficient as a function of time (see figure 3 of
Shalchi et al. 2011).

For $t \rightarrow \infty$, we only find a contribution to the $k_{\parallel}$-integral for the smallest possible
values of $\omega_{\pm}$ meaning $\omega_- = 0$ and $\omega_+ = - \kappa_{\parallel} k_{\parallel}^2$. Furthermore,
we derive from Eq. (\ref{omegapm})
\be
\omega_+ - \omega_- = 2 \sqrt{D^2 - \left( v k_{\parallel} \right)^2 / 3} \approx 2 D = v / \lambda_{\parallel}.
\ee
Therewith Eq. (\ref{dperpslab}) becomes
\be
d_{\perp} (t) = \frac{4 \pi \kappa_{\parallel}}{B_0^2} \int_{0}^{\infty} d k_{\parallel} \; g^{slab} \left( k_{\parallel} \right) e^{ - \kappa_{\parallel} k_{\parallel}^2 t}
\ee
where we have also used $\lambda_{\parallel} = 3 \kappa_{\parallel} / v$. This formula is in agreement with the result
originally obtained in Shalchi \& D\"oring (2007). As discussed there, the formula describes correctly compound sub-diffusion
as usually obtained for slab turbulence (see, e.g., K\'ota \& Jokipii 2000 and Webb et al. 2006). In the appendix of the current
paper we briefly show how quasi-linear theory can be recovered from Eq. (\ref{slabode}) as well.
\subsection{Diffusion Approximation}
Especially in non-linear treatments of particle transport it is often assumed that perpendicular transport is diffusive for
all times. In our notation this means that we set
\be
\langle ( \Delta x )^2 \rangle = 2 \kappa_{\perp} t \quad \forall \quad t
\label{diffapprox}
\ee
in Eq. (\ref{centraleq}). We like to emphasize that this is only an approximation. In reality one expects a ballistic motion
and thereafter there could be a sub-diffusive regime (see, e.g., Sects. 3.2 and 4 of the current paper). Eventually diffusion
is recovered if there is transverse structure (see, e.g., Shalchi 2017 and Sect. 4 of the current paper). It is the purpose
of the current article to explore the different transport regimes. In the current paragraph we employ the diffusion
approximation only to restore previous equations for the perpendicular diffusion coefficient. With approximation (\ref{diffapprox})
we derive from Eq. (\ref{centraleq})
\bdm
\langle V_x (t) V_x (0) \rangle & = & \frac{v^2}{3 B_0^2} \int d^3 k \; P_{xx} \left( \vec{k} \right) \frac{1}{\omega_+ - \omega_-} \nonumber\\
& \times & \left[ \omega_+ e^{\omega_+ t} - \omega_- e^{\omega_- t} \right] e^{- \kappa_{\perp} k_{\perp}^2 t}.
\edm
If we integrate the latter equation over time, and after employing the TGK formula (\ref{tgk}) for $t \rightarrow \infty$, we obtain
\bdm
\kappa_{\perp}
& = & \frac{v^2}{3 B_0^2} \int d^3 k \; P_{xx} \left( \vec{k} \right) \nonumber\\
& \times & \frac{\kappa_{\perp} k_{\perp}^2}{\left( \kappa_{\perp} k_{\perp}^2 \right)^2 - \left( \omega_- + \omega_+ \right) \kappa_{\perp} k_{\perp}^2 + \omega_- \omega_+}.
\edm
Now we use Eq. (\ref{omegapm}) to derive $\omega_- + \omega_+ = - 2 D = - v / \lambda_{\parallel}$ and
$\omega_- \omega_+ = (v k_{\parallel})^2 / 3$ to find
\be
\kappa_{\perp} = \frac{v^2}{3 B_0^2} \int d^3 k \;
\frac{P_{xx} (\vec{k})}{\kappa_{\perp} k_{\perp}^2 + v / \lambda_{\parallel} + F (k_{\parallel},k_{\perp})}
\label{UNLT}
\ee
where we have used the function
\be
F (k_{\parallel},k_{\perp}) = (v k_{\parallel})^2 /(3 \kappa_{\perp} k_{\perp}^2).
\label{defF}
\ee
This result agrees with the integral equation provided by UNLT theory (see Shalchi 2010) apart from a factor $4/3$
in front of the first term in the denominator of Eq. (\ref{UNLT}). The reason for the little discrepancy
is that Eq. (\ref{formforW}) itself is an approximation. One can easily repeat the calculations performed above for
the more general case of dynamical turbulence. In this case one would obtain the formula originally derived
in Shalchi (2011).
\section{Time-Dependent Transport in Turbulence with Transverse Complexity}
In the current section we study particle transport in different turbulence models with transverse structure. This is important
because the latter effect is essential in order to restore diffusion. As examples we consider the noisy slab model, a Gaussian
correlation model, pure two-dimensional turbulence, and a two-component turbulence model consisting of slab and two-dimensional
modes.
\subsection{Noisy Slab Turbulence}
As a first example for turbulence with transverse structure we consider the noisy slab model originally proposed in Shalchi (2015)
as a model with minimal transverse complexity. Within this model the magnetic correlation tensor has the components
\be
P_{nm} \left( \vec{k} \right) = \frac{2 \ell_{\perp}}{k_{\perp}} g^{slab} (k_{\parallel}) \Theta \left( 1 - k_{\perp} \ell_{\perp} \right)
\left( \delta_{nm} - \frac{k_n k_m}{k_{\perp}^2} \right)
\label{noisyslab}
\ee
where we have used the {\it Heaviside step function} $\Theta (x)$ and the perpendicular correlation scale of the
turbulence $\ell_{\perp}$. This form can be understood as broadened slab turbulence.  The latter model can be recovered
in the limit $\ell_{\perp} \rightarrow \infty$.

For the noisy slab model Eq. (\ref{centraleq}) becomes
\bdm
\langle V_x (t) V_x (0) \rangle & = & \frac{4 \pi v^2 \ell_{\perp}}{3 B_0^2}
\int_{0}^{\infty} d k_{\parallel} \; g^{slab} (k_{\parallel}) \nonumber\\
& \times & \frac{1}{\omega_+ - \omega_-} \left[ \omega_+ e^{\omega_+ t} - \omega_- e^{\omega_- t} \right] \nonumber\\
& \times & \int_{0}^{1/\ell_{\perp}} d k_{\perp} \; e^{- \frac{1}{2} \langle (\Delta x)^2 \rangle k_{\perp}^2}.
\edm
The perpendicular wavenumber integral therein can be expressed by an {\it error function} and we derive
\bdm
& & \frac{d^2}{d t^2} \langle (\Delta x)^2 \rangle \nonumber\\
& = & \frac{8 \pi v^2 \ell_{\perp}}{3 B_0^2} \sqrt{\frac{\pi}{2 \langle (\Delta x)^2 \rangle}}
\textnormal{Erf} \left( \sqrt{\frac{\langle (\Delta x)^2 \rangle}{2 \ell_{\perp}^2}} \right) \nonumber\\
& \times & \int_{0}^{\infty} d k_{\parallel} \; g^{slab} (k_{\parallel}) \frac{1}{\omega_+ - \omega_-}
\left[ \omega_+ e^{\omega_+ t} - \omega_- e^{\omega_- t} \right]. \nonumber\\
\label{Sigmanoisy2}
\edm
For a numerical evaluation of Eq. (\ref{Sigmanoisy2}), it is convenient to employ the integral transformation
$x=\ell_{\parallel} k_{\parallel}$, to use the Kubo number $K = (\ell_{\parallel} \delta B_x)/(\ell_{\perp} B_0)$,
the dimensionless time $\tau = \kappa_{\parallel} t / \ell_{\parallel}^2$, as well as
$\sigma = \langle (\Delta x)^2 \rangle / \ell_{\perp}^2$. 

\begin{figure}
\includegraphics[width=.48\textwidth]{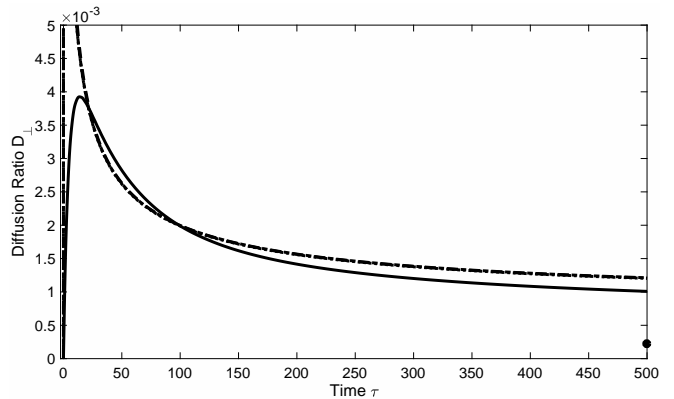}
\caption{The running diffusion ratio $D_{\perp} := (\ell_{\parallel}^2 d_{\perp})/(\ell_{\perp}^2 \kappa_{\parallel})$
versus time $\tau = \kappa_{\parallel} t / \ell_{\parallel}^2$ for the noisy slab model as obtained by solving Eq. (\ref{Sigmanoisy3})
numerically for a Kubo number of $K=0.2$. We have shown the results obtained for different values of the parallel mean free path,
namely $\lambda_{\parallel} / \ell_{\parallel} = 0.01$ (dotted line), $\lambda_{\parallel} / \ell_{\parallel} = 0.1$ (dash-dotted line),
$\lambda_{\parallel} / \ell_{\parallel} = 1$ (dashed line), as well as $\lambda_{\parallel} / \ell_{\parallel} = 10$ (solid line).
Note that dotted, dash-dotted, and dashed lines are in coincidence. The dot represents the result obtained by employing diffusive
UNLT theory.}
\label{NoisyslabK0p2}
\end{figure}

\begin{figure}
\includegraphics[width=.48\textwidth]{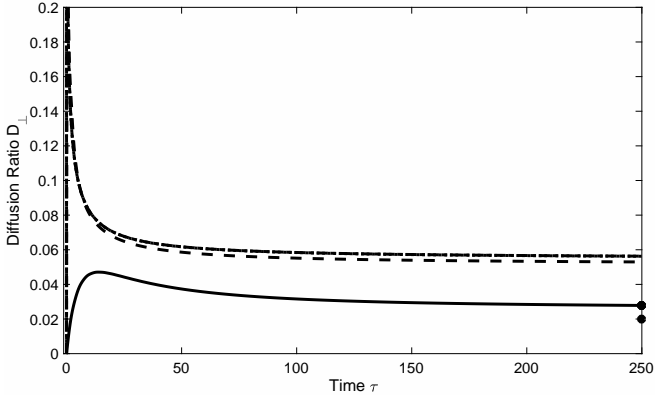}
\caption{Caption is as in Fig. \ref{NoisyslabK0p2} but here we have used $K=0.7$ for the Kubo number. Note that dotted and dash-dotted
lines are in coincidence.}
\label{NoisyslabK0p7}
\end{figure}

For the spectrum $g^{slab} (k_{\parallel})$ we use the Bieber et al. (1994) model
\be
g^{slab} (k_{\parallel}) = \frac{1}{2 \pi} C(s) \delta B^2 \ell_{\parallel} \left[ 1 + (k_{\parallel} \ell_{\parallel})^2 \right]^{-s/2}.
\label{slabspec}
\ee
Here we have used the normalization function
\be
C(s) = \frac{\Gamma \left( \frac{s}{2} \right)}{2 \sqrt{\pi} \Gamma \left( \frac{s-1}{2} \right)}
\label{normalC}
\ee
with the inertial range spectral index $s$ and gamma functions. For the spectral index in the inertial range we use $s=5/3$
throughout the whole paper as motivated by the famous work of Kolmogorov (1941). The parameter $\ell_{\parallel}$ is the bendover
scale in the parallel direction. After the integral transformation $z=k_{\parallel} \ell_{\parallel}$, Eq. (\ref{Sigmanoisy2})
can be written as
\bdm
\frac{d^2}{d \tau^2} \sigma
& = & 12 \sqrt{\pi} C(s) K^2 \frac{\ell_{\parallel}^2}{\lambda_{\parallel}^2} \sqrt{\frac{2}{\sigma}} \;
\textnormal{Erf} \left( \sqrt{\frac{\sigma}{2}} \right) \nonumber\\
& \times & \int_{0}^{\infty} d z \; \left( 1 + z^2 \right)^{-s/2} \nonumber\\
& \times & \frac{1}{\Omega_+ - \Omega_-} \left[ \Omega_+ e^{\Omega_+ \tau} - \Omega_- e^{\Omega_- \tau} \right]
\label{Sigmanoisy3}
\edm
where we have also used
\be
\Omega_{\pm} = \frac{\ell_{\parallel}^2}{\kappa_{\parallel}} \omega_{\pm}
= \frac{3 \ell_{\parallel}}{2 \lambda_{\parallel}}
\left[ - \frac{\ell_{\parallel}}{\lambda_{\parallel}} \pm \sqrt{\frac{\ell_{\parallel}^2}{\lambda_{\parallel}^2} - \frac{4}{3} z^2} \right].
\label{capomega}
\ee
Differential equation (\ref{Sigmanoisy3}) can be solved numerically. The corresponding running diffusion ratio\footnote{We
like to point out that there is a typo in Shalchi (2017). In the latter paper the definition
$D_{\perp} := (\ell_{\perp}^2 d_{\perp})/(\ell_{\parallel}^2 \kappa_{\parallel})$ was used. The way how the parameter $D_{\perp}$
is defined in the current article is correct.} $D_{\perp} := (\ell_{\parallel}^2 d_{\perp})/(\ell_{\perp}^2 \kappa_{\parallel})$
is shown in Figs. \ref{NoisyslabK0p2} and \ref{NoisyslabK0p7} for two different Kubo numbers. For the initial conditions we have set
$\sigma (0) = 0$ and $(d \sigma) / (d \tau) (0) = 0$ corresponding to a ballistic motion. This is used for all computations presented
in the current paper. In both considered cases we find a sub-diffusive motion directly after the ballistic regime. For a small Kubo number
the sub-diffusive regime persists for a long time whereas for an intermediate Kubo number diffusion is restored earlier.

\subsection{The Gaussian Correlation Model}
\begin{figure}
\includegraphics[width=.48\textwidth]{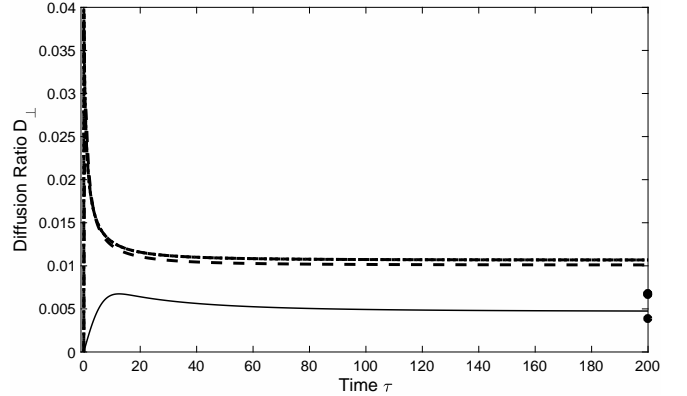}
\caption{The running diffusion ratio $D_{\perp} := (\ell_{\parallel}^2 d_{\perp})/(\ell_{\perp}^2 \kappa_{\parallel})$
versus time $\tau = \kappa_{\parallel} t / \ell_{\parallel}^2$ for the Gaussian model as obtained by solving Eq. (\ref{odegaussian3})
numerically for a Kubo number of $K=0.2$. We have shown the results obtained for different values of the parallel mean free path,
namely $\lambda_{\parallel} / \ell_{\parallel} = 0.01$ (dotted line), $\lambda_{\parallel} / \ell_{\parallel} = 0.1$ (dash-dotted line),
$\lambda_{\parallel} / \ell_{\parallel} = 1$ (dashed line), as well as $\lambda_{\parallel} / \ell_{\parallel} = 10$ (solid line).
Note that dotted and dash-dotted lines are in coincidence. The dots represent the corresponding solution of UNLT theory within the
diffusion approximation.}
\label{GaussianK0p2}
\end{figure}

\begin{figure}
\includegraphics[width=.48\textwidth]{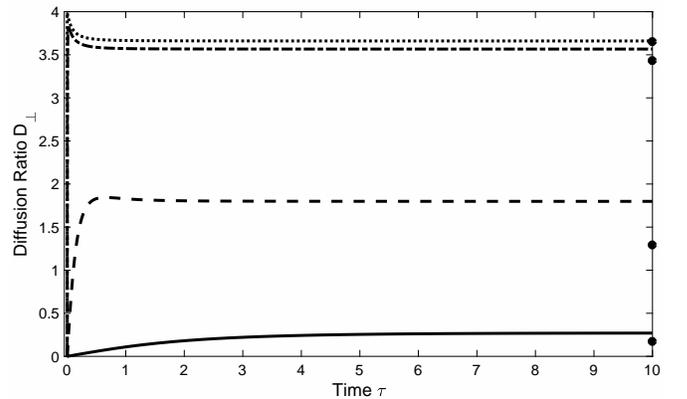}
\caption{Caption is as in Fig. \ref{GaussianK0p2} but here we have used $K=2$ for the Kubo number.\\[0.5cm]}
\label{GaussianK2}
\end{figure}

In the current subsection we employ a Gaussian correlation model which is often used as an example
(see, e.g., Neuer \& Spatschek 2006). In this case the components of the magnetic correlation tensor are given by
\bdm
P_{nm} (\vec{k}) & = & \frac{\ell_{\parallel} \ell_{\perp}^{4} \delta B_x^2}{(2 \pi)^{3/2}} k_{\perp}^2
e^{-\frac{1}{2} (\ell_{\parallel} k_{\parallel})^2 -\frac{1}{2} (\ell_{\perp} k_{\perp})^2} \nonumber\\
& \times & \left( \delta_{nm} - \frac{k_n k_m}{k_{\perp}^2} \right).
\label{neuerspek}
\edm
The parameters used in this model are the same as used above. If this model is combined with Eq. (\ref{centraleq2})
we derive
\bdm
\frac{d^2}{d t^2} \langle (\Delta x)^2 \rangle
& = & \frac{2 \ell_{\parallel} \ell_{\perp}^4}{\sqrt{2 \pi}} \frac{\delta B_x^2}{B_0^2} \nonumber\\
& \times & \int_{0}^{\infty} d k_{\parallel} \; \xi \left( k_{\parallel}, t \right) e^{-\frac{1}{2} (\ell_{\parallel} k_{\parallel})^2} \nonumber\\
& \times & \int_{0}^{\infty} d k_{\perp} \; k_{\perp}^3 e^{- \frac{1}{2} \left[ \ell_{\perp}^2 + \langle (\Delta x)^2 \rangle \right] k_{\perp}^2}.
\label{odegaussian}
\edm
The perpendicular wave number integral can be solved by
\be
\int_{0}^{\infty} d k_{\perp} \; k_{\perp}^3 e^{- \frac{1}{2} \left[ \ell_{\perp}^2 + \langle (\Delta x)^2 \rangle \right] k_{\perp}^2}
= 2 \left[ \ell_{\perp}^2 + \langle (\Delta x)^2 \rangle \right]^{-2}.
\label{perpint}
\ee
Therewith, Eq. (\ref{odegaussian}) becomes
\bdm
\frac{d^2}{d t^2} \langle (\Delta x)^2 \rangle
& = & \sqrt{\frac{8}{\pi}} \ell_{\parallel} \ell_{\perp}^4 \frac{\delta B_x^2}{B_0^2} \left[ \ell_{\perp}^2 + \langle (\Delta x)^2 \rangle \right]^{-2} \nonumber\\
& \times & \int_{0}^{\infty} d k_{\parallel} \; \xi \left( k_{\parallel}, t \right) e^{-\frac{1}{2} (\ell_{\parallel} k_{\parallel})^2}.
\label{odegaussian2}
\edm
For the parallel correlation function $\xi (k_{\parallel},t)$ we employ again model (\ref{formforW}). Using the
integral transformation $x=\ell_{\parallel} k_{\parallel}$, the Kubo number $K = (\ell_{\parallel} \delta B_x)/(\ell_{\perp} B_0)$,
the dimensionless time $\tau = \kappa_{\parallel} t / \ell_{\parallel}^2$, as well as $\sigma = \langle (\Delta x)^2 \rangle / \ell_{\perp}^2$
yields
\bdm
\frac{d^2}{d \tau^2} \sigma
& = & 6 \sqrt{\frac{2}{\pi}} \frac{\ell_{\parallel}^2}{\lambda_{\parallel}^2} K^2 \left[ 1 + \sigma \right]^{-2} \nonumber\\
& \times & \int_{0}^{\infty} d z \; \frac{1}{\Omega_+ - \Omega_-} \left[ \Omega_+ e^{\Omega_+ \tau} - \Omega_- e^{\Omega_- \tau} \right]
e^{-\frac{1}{2} z^2} \nonumber\\
\label{odegaussian3}
\edm
where the parameters $\Omega_{\pm}$ are given again by Eq. (\ref{capomega}). The latter differential equation can be solved
numerically. The results are visualized in Figs. \ref{GaussianK0p2} and \ref{GaussianK2} for two different values of the Kubo number.
As before we find ballistic, sub-diffusive, and normal diffusive regimes. For the Gaussian model, however, the sub-diffusive regime
is very short in particular for the case of large Kubo numbers.

\subsection{Two-Dimensional Turbulence}
\begin{figure}
\includegraphics[width=.48\textwidth]{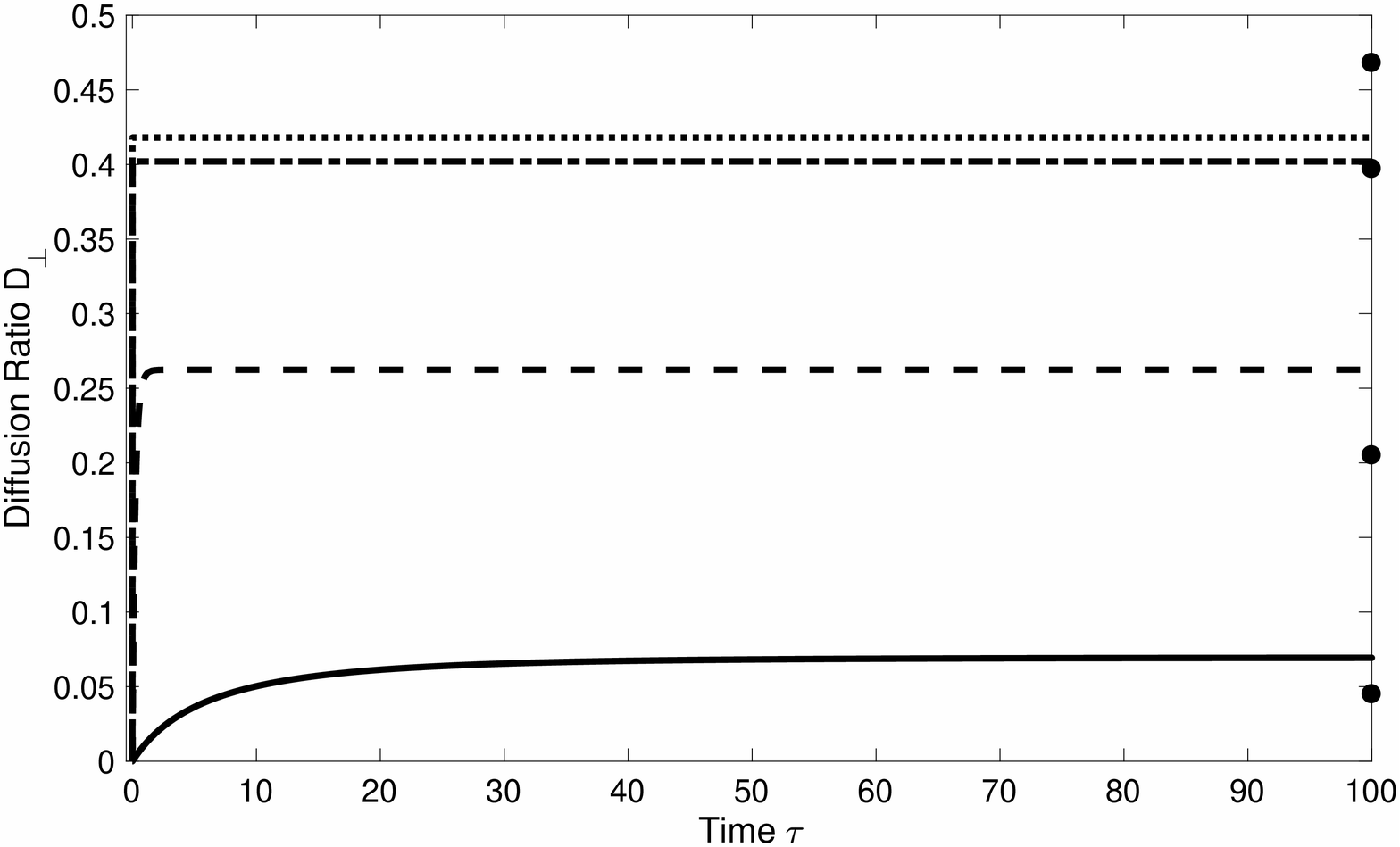}
\caption{The running diffusion ratio $D_{\perp} := (\ell_{\parallel}^2 d_{\perp})/(\ell_{\perp}^2 \kappa_{\parallel})$
versus time $\tau = \kappa_{\parallel} t / \ell_{\parallel}^2$ for two-dimensional turbulence as obtained by solving Eq. (\ref{2dode2})
numerically for a Kubo number of $K=0.7$ and $q=3$. We have shown the results obtained for different values of the parallel mean free path,
namely $\lambda_{\parallel} / \ell_{\parallel} = 0.01$ (dotted line), $\lambda_{\parallel} / \ell_{\parallel} = 0.1$ (dash-dotted line),
$\lambda_{\parallel} / \ell_{\parallel} = 1$ (dashed line), as well as $\lambda_{\parallel} / \ell_{\parallel} = 10$ (solid line).
The dots represent the result obtained by employing diffusive UNLT theory.}
\label{TwodimensionalK0p7}
\end{figure}

In the current paragraph we employ the so-called two-dimensional model where we have by definition
\be
P_{nm}^{2D} (\vec{k}) = g^{2D} (k_{\perp}) \frac{\delta (k_{\parallel})}{k_{\perp}} \left( \delta_{nm} - \frac{k_n k_m}{k_{\perp}^2} \right)
\label{Plm2D}
\ee
if $l,m=x,y$ and $P_{lz}=P_{zm}=P_{zz}=0$. In this particular model the magnetic field vector as well as the spatial
dependence are two-dimensional. Above we have used the spectrum of the two-dimensional modes $g^{2D} (k_{\perp})$
for which we employ the Shalchi \& Weinhorst (2009) model
\be
g^{2D}(k_{\perp}) = \frac{2 D(s, q)}{\pi} \delta B_{2D}^2 \ell_{\perp}
\frac{(k_{\perp} \ell_{\perp})^{q}}{\left[ 1 + (k_{\perp} \ell_{\perp})^2 \right]^{(s+q)/2}}.
\label{2dspec}
\ee
The latter spectrum contains a characteristic scale $\ell_{\perp}$ denoting the turnover from the energy range to the inertial
range. In the inertial range the spectrum scales like $k_{\perp}^{-s}$ whereas in the energy range it scales like $k_{\perp}^{q}$.
The energy range spectral index $q$ was discussed in detail in Matthaeus et al. (2007). In Eq. (\ref{2dspec}) we have used the
normalization function
\be
D(s, q) = \frac{\Gamma \left( \frac{s+q}{2} \right)}{2 \Gamma \left( \frac{s-1}{2} \right) \Gamma \left( \frac{q+1}{2} \right)}.
\label{normalD}
\ee
With the two-dimensional correlation tensor (\ref{Plm2D}), differential equation (\ref{centraleq2}) becomes
\bdm
\frac{d^2}{d t^2} \langle (\Delta x)^2 \rangle
& = & \frac{2 \pi}{B_0^2} \int_{0}^{\infty} d k_{\perp} \; g^{2D} \left( k_{\perp} \right) \nonumber\\
& \times & \xi \left( k_{\parallel} = 0, t \right) e^{- \frac{1}{2} \langle ( \Delta x )^2 \rangle k_{\perp}^2}.
\edm
From Eq. (\ref{formforW}) we deduce
\be
\xi \left( k_{\parallel}=0, t \right) = \frac{v^2}{3} e^{- v t / \lambda_{\parallel}}
\ee
and, thus, we derive
\bdm
\frac{d^2}{d t^2} \langle (\Delta x)^2 \rangle
& = & \frac{2 \pi v^2}{3 B_0^2} \int_{0}^{\infty} d k_{\perp} \; g^{2D} \left( k_{\perp} \right) \nonumber\\
& \times & e^{- v t / \lambda_{\parallel} - \frac{1}{2} \langle ( \Delta x )^2 \rangle k_{\perp}^2}.
\label{2dodewithspec}
\edm
To continue we use spectrum (\ref{2dspec}) and employ the integral transformation $z = k_{\perp} \ell_{\perp}$ to obtain
\bdm
\frac{d^2}{d t^2} \langle (\Delta x)^2 \rangle
& = & \frac{4}{3} D(s, q) v^2 \frac{\delta B_{2D}^2}{B_0^2} e^{- v t / \lambda_{\parallel}} \nonumber\\
& \times & \int_{0}^{\infty} d z \; \frac{z^q}{\left[ 1 + z^2 \right]^{(s+q)/2}} \nonumber\\
& \times & e^{- \frac{1}{2} \langle ( \Delta x )^2 \rangle z^2 / \ell_{\perp}^2}.
\label{2dode}
\edm
The latter integral can be expressed by {\it Tricomi's confluent hypergeometric function} (see, e.g., Gradshteyn \& Ryzhik 2000)
but this is not useful for our numerical investigations.
%
Using again the Kubo number $K = (\ell_{\parallel} \delta B_x)/(\ell_{\perp} B_0)$, the dimensionless
time $\tau = \kappa_{\parallel} t / \ell_{\parallel}^2$, as well as $\sigma = \langle (\Delta x)^2 \rangle / \ell_{\perp}^2$ yields
\bdm
\frac{d^2}{d \tau^2} \sigma
& = & 24 D(s, q) \frac{\ell_{\parallel}^2}{\lambda_{\parallel}^2} K^2 e^{- 3 \tau \ell_{\parallel}^2 / \lambda_{\parallel}^2} \nonumber\\
& \times & \int_{0}^{\infty} d z \; \frac{z^q}{\left[ 1 + z^2 \right]^{(s+q)/2}} e^{- \sigma z^2 / 2}.
\label{2dode2}
\edm
In the following the latter equation is solved numerically for an energy range spectral index of $q=3$. This value
is in agreement with the conditions discussed in Matthaeus et al. (2007). The result is visualized in Fig. \ref{TwodimensionalK0p7}.
In this case we find that diffusion is restored directly after the ballistic regime. A sub-diffusive regime cannot be observed.
We made some calculations for other values of the parameter $q$ but no qualitative difference was found.

\subsection{Two-Component Turbulence}
Above we have considered the slab model and the two-dimensional model as examples. It is often assumed that turbulence
in the solar wind can be approximated by a two-component model in which we consider a superposition of slab and
two-dimensional modes (see, e.g., Matthaeus et al. 1990, Zank \& Matthaeus 1993, Oughton et al. 1994, Bieber et al. 1996,
Matthaeus et al. 1996, Dasso et al. 2005, Shaikh \& Zank 2007, Hunana \& Zank 2010, and Zank et al. 2017).

In this case time-dependent UNLT provides the following differential equation
\bdm
& & \frac{d^2}{d t^2} \langle (\Delta x)^2 \rangle \nonumber\\
& = & \frac{8 \pi v^2}{3 B_0^2} \int_{0}^{\infty} d k_{\parallel} \; g^{slab} \left( k_{\parallel} \right) \frac{1}{\omega_+ - \omega_-}
\left[ \omega_+ e^{\omega_+ t} - \omega_- e^{\omega_- t} \right] \nonumber\\
& + & \frac{2 \pi v^2}{3 B_0^2} \int_{0}^{\infty} d k_{\perp} \; g^{2D} \left( k_{\perp} \right)
e^{- v t / \lambda_{\parallel} - \frac{1}{2} \langle ( \Delta x )^2 \rangle k_{\perp}^2}
\edm
where we have combined Eqs. (\ref{formforW}), (\ref{slabode}), and (\ref{2dodewithspec}). For the two spectra we employ
Eqs. (\ref{slabspec}) and (\ref{2dspec}), respectively.

Using again the Kubo number $K = (\ell_{\parallel} \delta B_x)/(\ell_{\perp} B_0)$, the dimensionless time
$\tau = \kappa_{\parallel} t / \ell_{\parallel}^2$, $\sigma = \langle (\Delta x)^2 \rangle / \ell_{\perp}^2$,
as well as the integral transformations $y = k_{\perp} \ell_{\perp}$ and $z = k_{\parallel} \ell_{\parallel}$ yields
\bdm
\frac{d^2}{d \tau^2} \sigma
& = & 24 \frac{\ell_{\parallel}^2}{\lambda_{\parallel}^2} K^2 \left[ C(s) \frac{\delta B_{slab}^2}{\delta B^2}
\int_{0}^{\infty} d z \; \left( 1 + z^2 \right)^{-s/2} \right. \nonumber\\
& \times & \frac{1}{\Omega_+ - \Omega_-} \left[ \Omega_+ e^{\Omega_+ \tau} - \Omega_- e^{\Omega_- \tau} \right] \nonumber\\
& + & D (s,q) \frac{\delta B_{2D}^2}{\delta B^2} \int_{0}^{\infty} d y \; \frac{y^q}{\left[ 1 + y^2 \right]^{(s+q)/2}} \nonumber\\
& \times & \left. e^{- 3 \tau \ell_{\parallel}^2 / \lambda_{\parallel}^2 - \frac{1}{2} \sigma y^2} \right]
\label{slab2dode}
\edm
where we have used the total turbulent magnetic field $\delta B^2 = \delta B_{slab}^2 + \delta B_{2D}^2 = 2 \delta B_x^2$.
For the energy range spectral index in the two-dimensional spectrum we set again $q=3$. We like to point out that the
Kubo number is used here for convenience only and to make it easier to compare the results obtained for two-component
turbulence with the other results. Strictly speaking, the two-component model is a composition of models with Kubo number
$K=0$ and $K=\infty$.

\begin{figure}
\includegraphics[width=.48\textwidth]{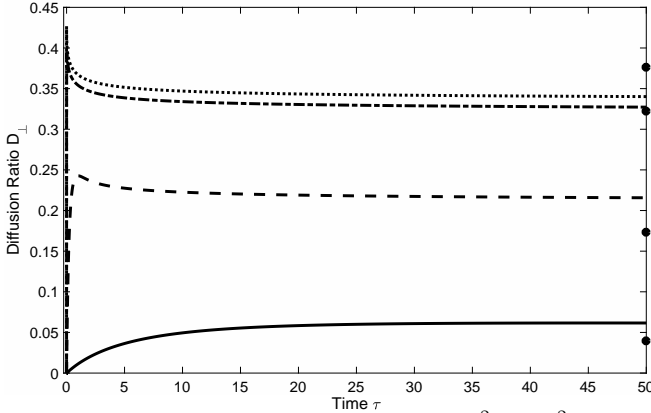}
\caption{The running diffusion ratio $D_{\perp} := (\ell_{\parallel}^2 d_{\perp})/(\ell_{\perp}^2 \kappa_{\parallel})$
versus time $\tau = \kappa_{\parallel} t / \ell_{\parallel}^2$ for slab/2D composite turbulence as obtained by solving Eq. (\ref{slab2dode})
numerically for a Kubo number of $K=0.7$. In this plot the slab fraction is $\delta B_{slab}^2 / \delta B^2 = 0.20$. We have shown the results
obtained for different values of the parallel mean free path, namely $\lambda_{\parallel} / \ell_{\parallel} = 0.01$ (dotted line),
$\lambda_{\parallel} / \ell_{\parallel} = 0.1$ (dash-dotted line), $\lambda_{\parallel} / \ell_{\parallel} = 1$ (dashed line), as well as
$\lambda_{\parallel} / \ell_{\parallel} = 10$ (solid line). The dots represent the result obtained by employing diffusive UNLT theory.}
\label{Slab2DK0p72080}
\end{figure}

\begin{figure}
\includegraphics[width=.48\textwidth]{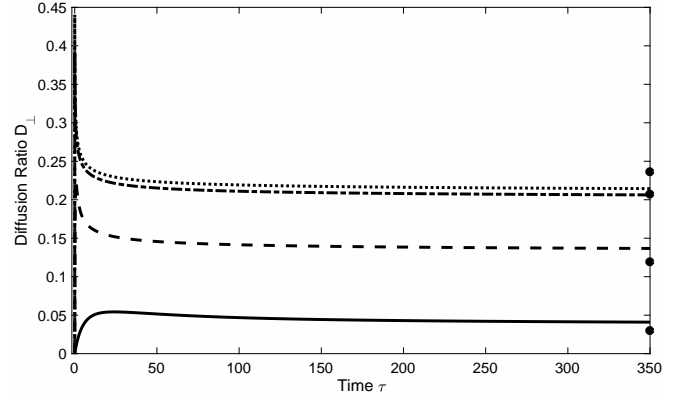}
\caption{Caption is as in Fig. \ref{Slab2DK0p72080} but here we have used $\delta B_{slab}^2 / \delta B^2 = 0.50$ for the
slab fraction.}
\label{Slab2DK0p75050}
\end{figure}

\begin{figure}
\includegraphics[width=.48\textwidth]{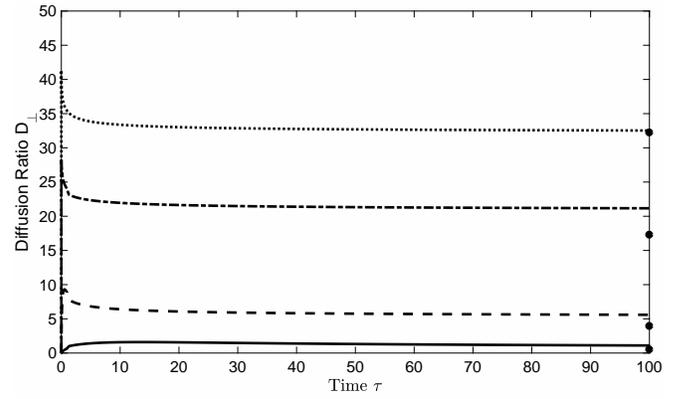}
\caption{Caption is as in Fig. \ref{Slab2DK0p72080} but here we have used $K=7$ for the Kubo number.}
\label{Slab2DK72080}
\end{figure}

In Figs. \ref{Slab2DK0p72080} and \ref{Slab2DK0p75050} we visualize the numerical solution of Eq. (\ref{slab2dode}) for the
case of an intermediate Kubo numbers of $K=0.7$. Furthermore, we have considered two values of the slab fraction, namely
$\delta B_{slab}^2/\delta B^2=0.2$ as originally suggested by Bieber et al. (1994) and Bieber et al. (1996) as well as
the balanced case $\delta B_{slab}^2/\delta B^2=0.5$. In Fig. \ref{Slab2DK72080} we show the results for a large Kubo
number of $K=7$ and a slab fraction of $\delta B_{slab}^2/\delta B^2=0.2$. The case is relevant because it corresponds
to a scale ration of $\ell_{\perp} = 0.1 \ell_{\parallel}$ used in some previous work (see, e.g., Matthaeus et al. 2003).
In all three cases we observe that for a long parallel mean free path the diffusive regime comes directly after the ballistic
regime. For shorter parallel mean free paths we find a ballistic regime, then sub-diffusion, and at later times diffusion
is recovered.

Important here is to note that the first contribution in Eq. (\ref{slab2dode}) corresponds to the usual slab result meaning
that it behaves sub-diffusively. However, the second term contains the mean square displacement $\sigma$. Therefore, we expect
an implicit contribution of the slab modes. The latter effect was already studied in Shalchi (2016). In the latter paper,
however, diffusion theory was extended by a sub-diffusive slab contribution. The equation used here is more general because
the parameter $\sigma$ is calculated without any assumption concerning diffusivity.

\section{Summary and Conclusion}
It is crucial in astrophysics, space science, and plasma physics to understand the motion of energetic particles across
a mean magnetic field. Previous analytical theories were based on the diffusion approximation together with a late time limit.
Therefore, such theories only provide formulas for the perpendicular diffusion coefficient or mean free path. Recently a
time-dependent version of the {\it unified non-linear transport (UNLT) theory} was presented in Shalchi (2017).
This theory is no longer based on the diffusion approximation and allows one to describe the transport for early 
times as well.

In the current paper we combined the aforementioned theory with a more general parallel transport model and different
turbulence models. We have considered transport in slab, noisy slab, Gaussian, two-dimensional, and two-component
turbulence. In all cases we have computed the running perpendicular diffusion coefficient as a function of time
in order to explore ballistic, sub-diffusive, and diffusive regimes.

For slab and noisy slab turbulence we find a sub-diffusive regime directly after the ballistic regime.
If transverse complexity is present, diffusion is restored as soon as the condition
$\langle (\Delta x)^2 \rangle \gg 2 \ell_{\perp}^2$ is satisfied. For all considered turbulence configurations we
find a similar form of the running diffusion coefficient. In some case we find normal diffusion directly after
the ballistic regime. This is in particular the case for large Kubo numbers and long parallel mean free paths.
In some other cases the sub-diffusive regime persists for a very long time before diffusion is restored. All our
results are compared with the results obtained by using standard UNLT theory involving a diffusion approximation.
We find that the diffusion approximation works well in the late time regime but a small discrepancy can be found
in some cases. This discrepancy could be relevant if analytical results are compared directly with test-particle
simulation and the diffusion approximation itself can contribute to the mysterious factor $a^2$ often used in non-linear
treatments of perpendicular diffusion (see, e.g., Matthaeus et al. 2003 and Shalchi 2010). It also has to be emphasized
that time-dependent UNLT theory also contains the effect of the implicit contribution of slab modes in two-component
turbulence as described in Shalchi (2016).

In the past authors discussed the possibility of non-diffusive transport in the literature (see, e.g., Zimbardo et al. 2006,
Pommois et al. 2007, Shalchi \& Kourakis 2007, and Zimbardo et al. 2012). The theory proposed in Shalchi (2017) and
used in the current paper allows, in principle, to describe sub- and super-diffusive transport. Indeed we found several
cases where the transport is sub-diffusive for a long time. However, in all considered cases Markovian diffusion is
restored in the late time limit if there is transverse complexity. If there is indeed non-diffusive transport in the late
time limit, it must either be for an extreme turbulence model or a non-diffusive parallel transport model. Such cases
could be explored in future work.

The theory used in the current paper and originally developed in Shalchi (2017) is the most advanced analytical theory
for perpendicular transport developed so far. It is able to describe perpendicular transport for arbitrary time and
transport behavior, including ballistic motion and compound sub-diffusion, but it is also a theory which is no longer
based on the diffusion approximation. The theory is still tractable because the perpendicular diffusion coefficient
can easily be computed by solving an ordinary differential equation numerically (see, e.g., Eq. (\ref{centraleq}) of the
current paper). One would, therefore, expect that there will be a variety of applications in astrophysics and space science.
\begin{acknowledgements}
{\it Support by the Natural Sciences and Engineering Research Council (NSERC) of Canada is acknowledged.}
\end{acknowledgements}
\appendix
\section{Quasi-Linear Perpendicular Diffusion}
The question arises how quasi-linear theory can be recovered from the time-dependent approach used in the
current paper. We assume slab turbulence as it was originally done in the work of Jokipii (1966). Therefore,
we start our investigations with Eq. (\ref{slabode}). In order to obtain the quasi-linear limit we have to
consider the formal limit $\lambda_{\parallel} \rightarrow \infty$ and Eq. (\ref{formforW}) becomes
\be
\xi \left( k_{\parallel}, t \right) = \frac{v^2}{3} \cos{\left( \frac{1}{\sqrt{3}} v k_{\parallel} t \right)}.
\ee
Using this in Eq. (\ref{slabode}) yields
\be
\frac{d^2}{d t^2} \langle (\Delta x)^2 \rangle
= \frac{8 \pi v^2}{3 B_0^2} \int_{0}^{\infty} d k_{\parallel} \; g^{slab} \left( k_{\parallel} \right)
\cos{\left( \frac{1}{\sqrt{3}} v k_{\parallel} t \right)}.
\ee
After integrating the latter formula over time and employing Eq. (\ref{rundiff}), we derive
\be
\kappa_{\perp} = \frac{4 \pi v^2}{3 B_0^2} \int_{0}^{\infty} d k_{\parallel} \; g^{slab} \left( k_{\parallel} \right)
\int_{0}^{\infty} d t \; \cos{\left( \frac{1}{\sqrt{3}} v k_{\parallel} t \right)}.
\ee
With the relations (see, e.g., Zwillinger 2012)
\be
\int_{0}^{\infty} d t \; \cos{\left( \alpha t \right)} = \pi \delta \left( \alpha \right)
\ee
and
\be
\delta \left( \alpha z \right) = \frac{1}{\left| \alpha \right|} \delta \left( z \right)
\ee
we finally obtain for the perpendicular mean free path
\be
\lambda_{\perp} = \frac{3}{v} \kappa_{\perp} = \frac{2 \sqrt{3} \pi^2}{B_0^2} g^{slab} \left( k_{\parallel}=0 \right)
\ee
which, apart from a factor $2 / \sqrt{3} \approx 1.15$, agrees with the well-known quasi-linear
formula (see, e.g., Shalchi 2005). The reason for this small disagreement is the fact that Eq. (\ref{formforW})
itself is an approximation.
{}


\begin{thebibliography}{}

\bibitem[Bieber et al.(1994)]{bieber94}
Bieber, J. W., Matthaeus, W. H., Smith, C. W., Wanner, W., Kallenrode, M.-B., \& Wibberenz, G. 1994, ApJ, 420, 294

\bibitem[Bieber et al.(1996)]{bieber96}
Bieber, J. W., Wanner, W., \& Matthaeus, W. H. 1996, JGR, 101, 2511

\bibitem[Corrsin(1959)]{cor59}
Corrsin, S. 1959, in Atmospheric Diffusion and Air Pollution, Advances in Geophysics, Vol. 6,
ed. F. Frenkiel \& P. Sheppard (New York: Academic), 161

\bibitem[Dasso et al.(2005)]{dasso05}
Dasso, S., Milano, L., Matthaeus, W. H., \& Smith, C. 2005, ApJ, 635, L181

\bibitem[Gradshteyn \& Ryzhik(2000)]{GradshteynRyzhik00}
Gradshteyn, I .S., \& Ryzhik, I. M. 2000. Table of integrals, series, and products (New York: Academic Press)

\bibitem[Green(1951)]{green51}
Green, M. S. 1951, JChPh, 19, 1036

\bibitem[Hunana \& Zank(2010)]{Hunana2010}
Hunana, P., \& Zank, G. P. 2010, ApJ, 718, 148

\bibitem[Jokipii(1966)]{jok66}
Jokipii, J. R. 1966, ApJ, 146, 480

\bibitem[K\'ota \& Jokipii(2000)]{KotaJokipii2000}
K\'ota, J., \& Jokipii, J. R. 2000, ApJ, 531, 1067

\bibitem[Kubo(1957)]{kubo57}
Kubo, R. 1957, JPSJ, 12, 570

\bibitem[Kolmogorov(1941)]{kol41}
Kolmogorov, A. N. 1941, Dokl. Akad. Nauk. SSSR, 30, 301

\bibitem[Matthaeus et al.(1990)]{matt90}
Matthaeus, W. H., Goldstein, M. L., \& Aaron, R. D. 1990, JGR, 95, 20673

\bibitem[Matthaeus et al.(1995)]{matt95}
Matthaeus, M. W., Gray, P. C., Pontius Jr., D. H., \& Bieber, J. W. 1995, PhRvL, 75, 2136

\bibitem[Matthaeus et al.(1996)]{matt96}
Matthaeus, W. H., Ghosh, S., Oughton, S., \& Roberts, D. 1996, JGR, 101, 7619

\bibitem[Matthaeus et al.(2003)]{matt03}
Matthaeus, W. H., Qin, G., Bieber, J. W., \& Zank, G. P. 2003, ApJ, 590, L53

\bibitem[Matthaeus et al.(2007)]{matt07}
Matthaeus, W. H., Bieber, J. W., Ruffolo, D., Chuychai, P., \& Minnie, J. 2007, ApJ, 667, 956

\bibitem[Neuer \& Spatschek(2006)]{NeuerSpatschek06}
Neuer, M., \& Spatschek, K. H. 2006, PRE, 73, 26404

\bibitem[Oughton et al.(1994)]{ought94}
Oughton, S., Priest, E., \& Matthaeus, W. H. 1994, J. Fluid. Mech., 280, 95

\bibitem[Pommois et al.(2007)]{Pomm2007}
Pommois, P., Zimbardo, G., \& Veltri, P. 2007, PhPl, 14, 012311

\bibitem[Qin et al.(2002a)]{qin2002a}
Qin, G., Matthaeus, W. H., \& Bieber, J. W. 2002a, GeoRL, 29, 1048

\bibitem[Qin et al.(2002b)]{qin02b}
Qin, G., Matthaeus, W. H., \& Bieber, J. W., 2002b, ApJ, 578, L117


\bibitem[Rechester \& Rosenbluth(1978)]{rechrosen78}
Rechester, A. B., \& Rosenbluth, M. N. 1978, PhRvL, 40, 38

\bibitem[Schlickeiser(2002)]{Schlickeiser2002}
Schlickeiser R. 2002, Cosmic Ray Astrophysics (Berlin: Springer)

\bibitem[Shaikh \& Zank(2007)]{shaik07}
Shaikh, D., \& Zank, G. P. 2007, ApJ, 656, L17

\bibitem[Shalchi(2005)]{shalchi05jgr}
Shalchi, A. 2005, JGR, 110, A09103

\bibitem[Shalchi(2006)]{shalchi06}
Shalchi, A. 2006, A \& A, 448, 809

\bibitem[Shalchi \& D\"oring(2007)]{shaldo07}
Shalchi, A., \&  D\"oring, H. 2007, Journal of Physics G: Nuclear and Particle Physics, 34, 859

\bibitem[Shalchi \& Kourakis(2007)]{shalkou07}
Shalchi, A., \& Kourakis, I. 2007, A \& A, 470, 405

\bibitem[Shalchi(2009)]{shal09book}
Shalchi, A. 2009, Nonlinear Cosmic Ray Diffusion Theories, Astrophysics and Space Science Library, Vol. 362 (Berlin: Springer)

\bibitem[Shalchi et al.(2009)]{Shalchietal2009}
Shalchi, A., \v{S}koda, T., Tautz, R. C., \& Schlickeiser, R. 2009, A\&A, 507, 589

\bibitem[Shalchi \& Weinhorst(2009)]{Shalwei2009}
Shalchi, A., \&  Weinhorst, B. 2009, AdSpR, 43, 1429

\bibitem[Shalchi(2010)]{shal2010}
Shalchi, A. 2010, ApJL, 720, L127

\bibitem[Shalchi(2011)]{shal2011}
Shalchi, A. 2011, PPCF, 53, 074010

\bibitem[Shalchi et al.(2011)]{shaletal2011}
Shalchi, A., Tautz, R. C., \& Rempel, T. J. 2011, PPCF, 53, 105016

\bibitem[Shalchi(2015)]{shal2015}
Shalchi, A. 2015, PhPl, 22, 010704

\bibitem[Shalchi(2016)]{shal2016}
Shalchi, A. 2016, ApJ, 830, 130

\bibitem[Shalchi(2017)]{shal2017}
Shalchi, A. 2017, PhPl, 24, 050702

\bibitem[Taylor(1922)]{taylor22}
Taylor, G. I. 1922, Proceedings of the London Mathematical Society, 20, 196

\bibitem[Webb et al.(2006)]{Webbetal06}
Webb, G. M., Zank G. P., Kaghashvili E. Kh., \& le Roux, J. A. 2006, ApJ, 651, 211

\bibitem[Zank \& Matthaeus(1993)]{zankmat93}
Zank, G. P., \& Matthaeus, W. H. 1993, Physics of Fluids A, 5, 257

\bibitem[Zank(2014)]{zank14}
Zank, G. P. 2014, {\it Transport Processes in Space Physics and Astrophysics: Lecture Notes in Physics, 877} (New York: Springer)

\bibitem[Zank et al.(2017)]{Zanketal17}
Zank, G. P., Adhikari, L., Hunana, P., Shiota, D., Bruno, R., \& Telloni, D. 2017, ApJ, 835, 147

\bibitem[Zimbardo et al.(2006)]{zimb06}
Zimbardo, G., Pommois, P., \& Veltri, P. 2006, ApJ, 639, L91

\bibitem[Zimbardo et al.(2012)]{zimb12}
Zimbardo, G., Perri, S., Pommois, P., \& Veltri, P. 2012, AdSpR, 49, 1633

\bibitem[Zwillinger(2012)]{zwillinger12}
Zwillinger, D. 2012, Standard Mathematical Tables and Formulae (Boca Raton: CRC Press)

\end{thebibliography}
\end{document}